\documentclass{PoS}
\usepackage{amsmath}  
\usepackage{amssymb} 
 \usepackage{multirow}
\usepackage{psfrag,scalefnt}

\newcommand{\VEV}[1]{\ensuremath{\left\langle #1\right\rangle}}
\newcommand{\VEVsmall}[1]{\ensuremath{\langle #1\rangle}}
\newlength{\textlength}
\newlength{\overlinelength}
\newcommand{\ovl}[2][.55]{\settowidth{\textlength}{$#2$}
  \setlength{\overlinelength}{0.1pt}
  \addtolength{\overlinelength}{0.75\textlength}
  \makebox[\textlength][s]{$#2$} \hspace{-.55\textlength}
  \hspace{-\overlinelength}\hspace{#1\overlinelength}
  \overline{\makebox[\overlinelength][s]{\vphantom{$#2$}}}
  \hspace{-#1\overlinelength}\hspace{.55\textlength}}
\newcommand{\bbs}{\ensuremath{B_s\!-\!\ovl{\!B}_s\,}}
\newcommand{\bbms}{\bbs\ mixing}
\newcommand{\bbd}{\ensuremath{B_d\!-\!\ovl{\!B}_d\,}}
\newcommand{\bbmd}{\bbd\ mixing}
\newcommand{\kk}{\ensuremath{K\!-\!\ovl{\!K}\,}}
\newcommand{\kkm}{\kk\ mixing}

\title{Flavour Physics in an SO(10) Grand Unified  Model}

\ShortTitle{Flavour Physics in an SO(10) GUT}

\author{\speaker{Jennifer Girrbach}\\
        Technical University of Munich (TUM)\\
	Institute of Advanced Study (IAS)\\
        E-mail: \email{jennifer.girrbach@ph.tum.de}}

\abstract{Grand unified theories open the possibility to transfer the neutrino mixing matrix $U_\text{PMNS}$ to the
quark sector. This is accomplished  in a controlled way in a supersymmetric grand-unified model  proposed by Chang, Masiero and
Murayama (CMM model) where the atmospheric neutrino mixing angle induces large new $b\to s$ and $\tau\to\mu$ transitions. 
Relating the supersymmetric low-energy parameters to seven new parameters $ a_0,\, m_0^2,\, m_{\tilde g},\, D,\, \xi,\, \tan\beta$
and $\arg(\mu)$ of this SO(10) model, we perform a correlated study of several flavour-changing neutral current (FCNC) processes. 
The CMM model can serve as an alternative benchmark scenario to the popular constraint MSSM.

}

\FullConference{The 2011 Europhysics Conference on High Energy Physics, EPS-HEP 2011,\\
		July 21-27, 2011\\
		Grenoble, Rh\^one-Alpes, France}

\begin{document}

\section{Introduction}

Supersymmetric grand unified theories (SUSY GUTs) are popular extensions of the Standard Model (SM).
The generic Minimal Supersymmetric Standard Model (MSSM) has (too) many sources of
flavour and CP violation which reside in the soft breaking terms.
Contrarily, in the minimal flavour violating (MFV)  version of the MSSM large effects in the flavour sector can only appear in
very few processes, such as $b\to s \gamma$. SUSY GUTs can lie somewhere in
between. The unification of quarks and leptons into symmetry multiplets implies additional relations between SM parameters and
 correlation between the flavour mixing. 
Let's consider SU(5) multiplets:
 \begin{align}
  \overline{\mathbf{5}}_1 = \begin{pmatrix}
                            d_R^c\\
 			    d_R^c\\
 			    d_R^c\\
 			    e_L\\
 			    -\nu_e
                            \end{pmatrix},\qquad
 \overline{\mathbf{5}}_2 = \begin{pmatrix}
                             s_R^c\\
 			    s_R^c\\
 			    s_R^c\\
 			    \mu_L\\
 			    -\nu_\mu
                            \end{pmatrix},\qquad
 \overline{\mathbf{5}}_3 = \begin{pmatrix}
                             b_R^c\\
 			    b_R^c\\
 			    b_R^c\\
 			    \tau_L\\
 			    -\nu_\tau
                            \end{pmatrix}.
 \end{align}
If the PMNS matrix $U_\text{PMNS}$ stems from a mixing of these $\mathbf{5}$-plets, then the corresponding mixing angles should
also occur in the
charged-lepton sector and  right-handed down-quark sector. Especially the large atmospheric neutrino mixing angle
$\theta_{23}\approx 45^\circ$ induces $b_R\to s_R$ and $\tau_L\to \mu_L$ transitions. Whereas mixing of right-handed quark fields
in flavour space is unphysical it is not for the corresponding superfields due to the soft breaking terms. Consequently
squark-gluino loops can induce $b_R\to s_R$ transitions. Further slepton-neutralino/sneutrino-chargino loops  can induce $\tau\to
\mu$
transitions at an observable level. This was the main idea of Moroi and Chang, Masiero and Murayama in
\cite{Moroi:2000tk,Chang:2002mq}. Similar and
related works can be found e.g. in \cite{Barbieri:1995rs,Harnik:2002vs}.
In \cite{Girrbach:2011an} we have performed a global analysis in the CMM model including
an extensive renormalization group (RG) analysis to connect Planck-scale and low-energy parameters. In the next section we sketch
the theoretical framework focusing on the flavour structure.

\section{The CMM model -- a new benchmark scenario}

\subsection{Theory}

The idea of  PMNS-like mixing of down-quark singlets and lepton doublets as discussed above is encoded in the following
SO(10) superpotential:
\begin{align}\label{equ:superpotential}
  W_Y^\text{SO(10)}  =  \frac{1}{2}\mathbf{16}_i\, \mathsf{Y}^{ij}_1\, \mathbf{16}_j\,
  \mathbf{10}_H +  \mathbf{16}_i\, \mathsf{Y}^{ij}_2\,
  \mathbf{16}_j\, \frac{\mathbf{45}_H\, \mathbf{10}_H^\prime}{2M_\text{Pl}}
 +
  \, \mathbf{16}_i\, \mathsf{Y}_N^{ij}\, \mathbf{16}_j
  \frac{\overline{\mathbf{16}}_H \overline{\mathbf{16}}_H}{2M_\text{Pl}},
\end{align}
where $M_\text{Pl}$ is the Planck mass, $\mathbf {16}_i$ ($i = 1,2,3$) is the SO(10) spinor representation (one matter field per
generation) and $\mathbf{10}_H$, $\mathbf{10}_H^\prime$,
$\mathbf{45}_H$ and $\overline{\mathbf{16}}_H$ are four Higgs superfields, where $\mathbf{10}_H$ contains the MSSM $H_u$ and
$\mathbf{10}_H^\prime$ the MSSM $H_d$. One assumption of the CMM
model is that $\mathsf{Y}_1$ and $ \mathsf{Y}_N$ are simultaneously diagonalisable 
which can be achieved through a suitable flavour
symmetry at $M_\text{Pl}$. This flavour symmetry is broken by the second term
in (\ref{equ:superpotential}) with the consequence that the rotation matrix of the
right-handed down-squarks is exactly $U_\text{PMNS}$. SUSY is broken flavour blind at $M_\text{Pl}$ 
implying universal soft- and trilinear terms. That is,
the nonrenormalisable term $\propto
\mathsf{Y}_2$ in the superpotential contains the
whole flavour structure, its diagonalisation involves the PMNS and CKM
matrices (up to rephasings).
 The symmetry breaking chain reads
\begin{align}\label{equ:breaking}
  \text{SO(10)} \xrightarrow[{\VEV{\text{45}_H}}]{\VEV{\text{16}_H},\VEV{\overline{\text{16}}_H}}
  \text{SU(5)} \xrightarrow{\VEV{\text{45}_H}}
   \text{G}_\text{SM} 
  \xrightarrow{\VEV{\text{10}_H},\, \VEVsmall{\text{10}_H^\prime}}
  \text{SU(3)}_C \times \text{U(1)}_\text{em}\,,
\end{align}
which gives naturally small $\tan\beta$. 
 Then $\mathsf{Y}_1$ gives masses to up-type fermions, $\mathsf{Y}_2$ to down-type fermions and $\mathsf{Y}_N$ to
right-handed Majorana neutrinos.  We want to
stress that flavour physics observables depend  very weakly on the details  of the Higgs potential which was not specified in
the original paper \cite{Chang:2002mq}. But our results motivate further work on the Higgs
potential.

The key ingredient for the flavour structure is the following: In a weak basis with diagonal up-type Yukawa matrix we
have
\begin{align}
\mathsf Y_d = \mathsf Y_\ell^\top =V_\text{CKM}^\star 
\begin{pmatrix}
 y_d & 0 & 0 \\
0 & y_s & 0 \\
0 &0 & y_b
\end{pmatrix}
U_D\,,\qquad   U_D = U_\text{PMNS}^\ast\,\text{diag}(1,\, e^{i\xi},\,1)
\end{align}
and the right-handed down squark mass matrix at the low scale reads
\begin{align}
 m_{\tilde{d}}^2(M_Z) =
\textrm{diag}\left(m_{\tilde{d}_1}^2,m_{\tilde{d}_1}^2,m_{\tilde{d}_1}^2
\left(1-\Delta_{\tilde{d}}\right)\right)\,,
\end{align}
where $\Delta_{\tilde{d}}\in [0,\,1]$ defines the relative mass splitting between the 1$^{st}$/2$^{nd}$ and 3$^{rd}$ down-squark
generation. It is generated by RG effects of the top Yukawa coupling and can easily reach $0.4$. If we rotate to mass eigenstate
basis and diagonalise $\mathsf Y_d$ the neutrino mixing enters $m_{\tilde{D}}^2$:
\begin{align}
 m_{\tilde{D}}^2 = U_D m_{\tilde{d}}^2 U_D^\dagger=
m_{\tilde{d}_1}^2\left(\begin{array}{ccc}
1 & 0 & 0\\
0 & 1-\frac{1}{2}\Delta_{\tilde{d}} &
-\frac{1}{2}\Delta_{\tilde{d}}e^{i\xi}\\
0 & -\frac{1}{2}\Delta_{\tilde{d}}e^{-i\xi}&
1-\frac{1}{2}\Delta_{\tilde{d}}
                                \end{array}
\right)\,.
\end{align}
Consequently, the 23-entry $ \propto\Delta_{\tilde{d}}$ is responsible for $\tilde b_R-\tilde s_R$-mixing and exactly here a new
CP phase $\xi$ enters that affects \bbms. Note that there are zeros in the 12- and 13-entries, thus no effects in \kk\ and \bbmd\
appear.
This is due to the degeneracy of the first two squark generation and the assumed tribimaximal structure of $U_\text{PMNS}$.

\subsection{Comparison with CMSSM/mSUGRA}

Only seven parameters of the CMM model are relevant for our
analysis: the universal scalar soft mass $m_0$ and trilinear coupling
$a_0$ at the Planck scale, the gluino mass $m_{\tilde g}$, the $D$-term mass splitting $D$, the phase of $\mu$, the phase $\xi$
and $\tan\beta$ (but $2.7\lesssim\tan\beta\lesssim 10$). We did a comprehensive RG evolution to relate Planck-scale inputs to a
set of low-energy inputs: the  masses of
$\tilde u_R$ and $\tilde d_R$ of the first generations $m_{\tilde u_1},\,m_{\tilde d_1}$,  the 11-element of the trilinear
coupling of the down squarks $a_1^d$, $m_{\tilde g}$, $\arg\mu$, $\xi$ and $\tan\beta$. We evolve these parameters  twice from
$M_\text{ew}$ to $M_\text{Planck}$ and back to $M_\text{ew}$  to find all particle masses and MSSM couplings. 

\begin{table}[!tbh]
\begin{center}
\begin{small}
  \begin{tabular}{|c|c|c|}
\hline
generic MSSM & mSUGRA/CMSSM & CMM model \\
\hline
 \hline
$\approx$ 120 parameters & 4 parameters \& 1 sign & 7 input parameters\\
\hline
 \multirow{2}{*}{SUSY flavour \& CP problem}& minimize flavour & \multirow{2}{*}{clear flavour structure}\\
& violation ad-hoc&\\
\hline 
 \multirow{2}{*}{no universality} &  \multirow{2}{*}{universality at $M_\text{GUT}$}& universality at $M_\text{Pl}$\\
&&but broken at $M_\text{GUT}$ \\
\hline
\multicolumn{2}{|c|}{quarks \& leptons unrelated}&quark-lepton-interplay\\
\hline
Problem: suppress large &cannot explain current& can fit $\phi_s$ and small \\
effects elsewhere&flavour data (e.g. $\phi_s$)&effects in 1st/2nd gen.\\\hline
  \end{tabular}
\end{small}
\end{center}
\caption{Comparison between the generic MSSM, mSUGRA/CMSSM and the CMM model}\label{tab:comparison}
\end{table}

The minimal supergravity (mSUGRA) scenario or its popular variant, the constraint MSSM (CMSSM), has~-- similar to the
CMM model~-- only a few input parameters. But the philosophy is somewhat different: the CMSSM minimises flavour violation in
an ad-hoc way and assumes flavour universality at the GUT scale  with quark and lepton flavour structures being
unrelated. However the CMM model has a clear flavour structure
different from MFV and universality is already broken at $M_\text{GUT}$. Furthermore  due to this free
phase $\xi$, one can fit the \bbms\ phase $\phi_s$ to the data. Also the particle spectrum is quite different between the CMSSM
and the CMM model (mainly due to the large
mass splitting $\Delta_{\tilde d}$). 
This comparison is summarized
in tab.~\ref{tab:comparison}.
 Hence the CMM model could serve as a new benchmark model: it is well-motivated, has only
seven input parameters and it is a very predictive alternative to the well-studied CMSSM.

\subsection{Phenomenology}

\begin{figure}[!tbh]
 \psfrag{amsq}{\hspace{0.3cm}\scalefont{1.5}$\frac{a_1^d}{M_{\tilde q}}$}
  \psfrag{msq}{\hspace{-0.5cm}\scalefont{1.1}$M_{\tilde q}$[GeV]}
  \psfrag{mg500argmu0tanb6}{\hspace{-1.8cm}\scalefont{1}$m_{\tilde g_3} = 500~$GeV, $ \text{sgn}(\mu) = +1$, $\tan\beta = 6$}
\begin{minipage}{0.5\textwidth}
 \includegraphics[width = 0.95\textwidth]{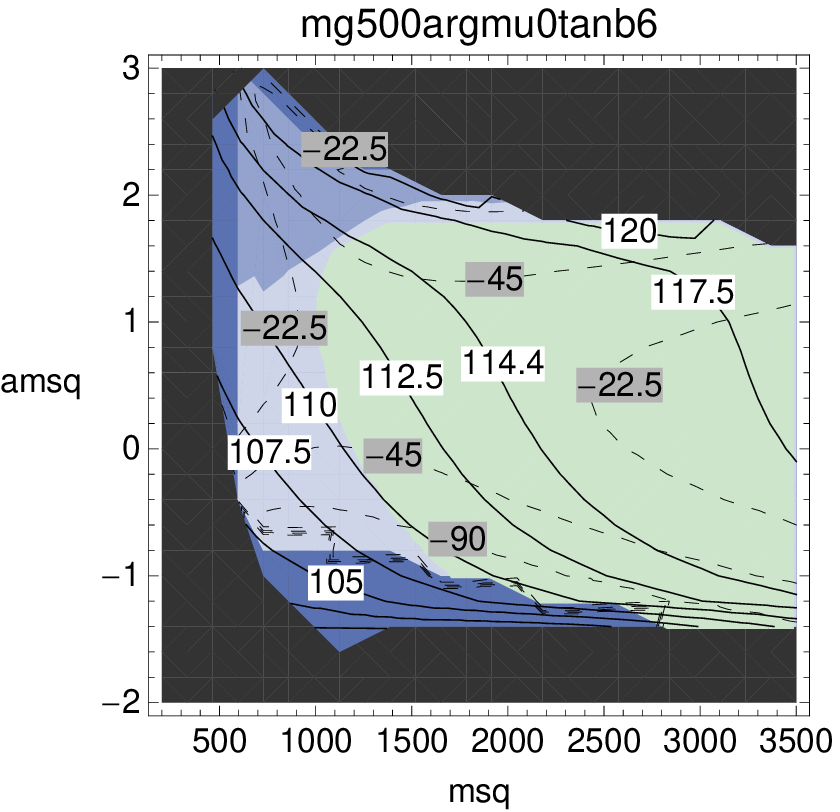}
\end{minipage}
\begin{minipage}{0.49\textwidth}
\begin{itemize}\setlength{\itemsep}{0pt}
 \item black: $m_{\tilde f}^2<0$, unstable vacuum
\item dark blue: excluded by \bbs
\item medium blue: excluded by $b\to s\gamma$
\item  light blue: excluded by $\tau\to \mu\gamma$
\item green: compatible with \bbs, $b\to s\gamma$, $\tau\to\mu\gamma$
\item Higgs mass in GeV: solid line with white labels
\item $\phi_s$ with maximal   possible
$|\phi_s|$ in degrees: dashed line with gray labels 
\end{itemize}

\end{minipage}
\caption{Correlation of FCNC processes as a function of $M_{\tilde q}(M_Z)$ (degenerate squark mass of first two generations) and
$a_1^d(M_Z)/M_{\tilde q}(M_Z)$ for $m_{\tilde g}(M_Z) = 500~\text{GeV}$ and $\text{sgn}(\mu) = +1$ with  $\tan\beta =6$.  }
\label{fig:result}
\end{figure}

We did a global analysis of flavour observables where we expected large CMM effects, namely \bbms, $b\to s  \gamma$ and $\tau\to
\mu\gamma$. Moreover we included vacuum stability bounds, lower bounds on sparticle masses and the mass of the lightest Higgs
boson. The result is shown in fig.~\ref{fig:result}. The flavour effects are proportional to $\Delta_{\tilde d}$ and maximized
for small $\tan\beta$. However, the Higgs mass constraint excludes too small values for $\tan\beta$. With $\xi$ we can
accommodate  a large $\phi_s$ while simultaneously fulfilling all other experimental constraints. The branching ratio
$BR(B_s\to\mu^+\mu^-)$ does not get large CMM effects because $\tan\beta$ is small. Realistic GUTs involve
dimension-5 Yukawa terms to fix the relation $\mathsf Y_d = \mathsf Y_\ell^\top$ for the 1$^{st}$ and 2$^{nd}$ generation.
Consequently we do not only get $b_R\to s_R$  but also $b_R\to d_R$ and $d_R\to s_R$ transition. This has been worked out in
\cite{Trine:2009ns} and is strongly constraint by \kkm. Similar constraints can be found from $\mu\to e\gamma$
\cite{Girrbach:2009uy}.

\section{Conclusion}

SUSY GUTs are theoretical well-motivated scenarios with correlations between hadronic and leptonic observables. If large CP
violation in \bbms\ is confirmed we need physics beyond the CMSSM and mSUGRA. We advertise the CMM model where the large
atmospheric neutrino mixing angle $\theta_{23}\approx 45^\circ$ induces $b-s$- and $\tau-\mu$-transitions as an alternative
benchmark scenario. We did an extensive RG analysis of the CMM model  relating several observables (\bbms, $b\to s\gamma$,
$\tau\to\mu\gamma$, $m_h$, vacuum stability bounds and lower bounds on sparticle masses) to
seven new input parameters beyond those of the SM. Due to a free phase $\xi$ we can adjust CP violation in \bbms\ while at the
same time getting only minor effects in $2\to 1$ and $3\to 1$ transitions.

\acknowledgments

I thank the organisers for the opportunity to give this talk and
my collaborators   S.~J\"ager, M.~Knopf, W.~Martens, U.~Nierste, C.~Scherrer and
S.~Wiesenfeldt for an enjoyable collaboration.  I thank U.~Nierste for proofreading this manuscript. My
work on the presented topic was supported by the \emph{Studienstiftung des
deutschen Volkes} and I acknowledge financial support by the DFG
cluster of excellence  ``Origin and Structure of the Universe''.


\begin{thebibliography}{99}


\bibitem{Moroi:2000tk}
  T.~Moroi,
  {JHEP {\bf 0003} (2000) 019}.
  [hep-ph/0002208];\\
    Phys.\ Lett.\  B {\bf 493} (2000) 366.
  [hep-ph/0007328].

  \bibitem{Chang:2002mq}
    D.~Chang, A.~Masiero and H.~Murayama,
    Phys.\ Rev.\  D {\bf 67} (2003) 075013,
  [hep-ph/0205111].


\bibitem{Barbieri:1995rs}
    R.~Barbieri, L.~J.~Hall and A.~Strumia,
    Nucl.\ Phys.\  B {\bf 449} (1995) 437.
[hep-ph/9504373];\\
  {Nucl.\ Phys.\  B {\bf 445} (1995) 219}.
[hep-ph/9501334].\\
%
   \bibitem{Harnik:2002vs}
    R.~Harnik, D.~T.~Larson, H.~Murayama and A.~Pierce,
    Phys.\ Rev.\  D {\bf 69} (2004) 094024. [hep-ph/9501334].


\bibitem{Girrbach:2011an}
  J.~Girrbach, S.~Jager, M.~Knopf, W.~Martens, U.~Nierste, C.~Scherrer and S.~Wiesenfeldt,
  JHEP {\bf 1106} (2011) 044
  [Erratum-ibid.\  {\bf 1107} (2011) 001]
  [arXiv:1101.6047 [hep-ph]].

\bibitem{Trine:2009ns}
  S.~Trine, S.~Westhoff, S.~Wiesenfeldt,
  JHEP {\bf 0908 } (2009)  002.
  [arXiv:0904.0378 [hep-ph]].

\bibitem{Girrbach:2009uy}
  J.~Girrbach, S.~Mertens, U.~Nierste, S.~Wiesenfeldt,
  JHEP {\bf 1005 } (2010)  026.
  [arXiv:0910.2663 [hep-ph]].
\\
  P.~Ko, J.~-h.~Park, M.~Yamaguchi,
  JHEP {\bf 0811 } (2008)  051.
  [arXiv:0809.2784 [hep-ph]].

\end{thebibliography}
\end{document}